\documentclass{article}
\usepackage[boxruled,vlined,linesnumbered]{algorithm2e}

\usepackage[english]{babel}
\usepackage[utf8]{inputenc}
\usepackage{fancyhdr}
\usepackage{graphicx,times,psfrag,xspace,color,latexsym,amsmath,amssymb}
\usepackage{comment}
\usepackage{slashbox}

\newcommand{\be}{\begin{eqnarray}}
\newcommand{\ee}{\end{eqnarray}}

\def\tK{\tilde{K}}

\def\ie{{\em i.e.},~}

\def\cf{{\em cf.}~}

\def \ux {{\underline{x}}}

\newcommand{\ben}{\begin{enumerate}}
\newcommand{\een}{\end{enumerate}}
\newcommand{\beq}{\begin{equation}}
\newcommand{\eeq}{\end{equation}}
\newcommand{\beqa}{\begin{eqnarray*}}
\newcommand{\eeqa}{\end{eqnarray*}}
\newcommand{\bit}{\begin{itemize}}
\newcommand{\eit}{\end{itemize}}
\newcommand{\bt}{\begin{tabular}{c}}
\newcommand{\btt}{\begin{tabular}}
\newcommand{\et}{\end{tabular}}

\newif\ifTR
\TRtrue

\newif\ifanon
\anonfalse

\begin{document}
\title{Online Scheduling of Spark Workloads with Mesos using Different Fair Allocation Algorithms\thanks{This work supported in part by an NSF CSR research grant, a gift from IBM, and a gift of AWS credits.}}


\author{
Y. Shan, A. Jain, G. Kesidis, B. Urgaonkar \\
School of EECS, PSU, State College, PA \\
\{yxs182,axj182,gik2,buu1\}@psu.edu  \\
~\\
 J. Khamse-Ashari and I. Lambadaris\\
SCE Dept, Carleton Univ., Ottawa, Canada\\
\{jalalkhamseashari,ioannis\}@sce.carleton.ca
}

\maketitle

\section{Introduction}

In the following, we present illustrative example and
experimental results comparing
fair schedulers 
allocating resources (indexed $r$) 
from multiple servers  (indexed $i$, with resource capacities $c_{i,r}$) to
distributed application frameworks (indexed $n$, with
resource demands per task $d_{n,r}$).
Resources are allocated  so that at least one resource $r$
is exhausted in every server.

Schedulers considered include DRF (DRFH) 
and Best-Fit DRF (BF-DRF) \cite{DRF,BLi15},
TSF \cite{BLi16b}, and PS-DSF \cite{Jalal17a}.
We also consider server selection under Randomized Round Robin (RRR)
and based on their residual (unreserved) resources.
In the following, we consider cases with frameworks of equal priority and 
without server-preference constraints.
We first give typical results of an illustrative numerical study and then give typical results of a study involving Spark workloads on Mesos, which we have modified and open-sourced to prototype different schedulers.

\section{Illustrative numerical study of 
fair scheduling by progressive filling}\label{sec:numer}



In this section, we 
consider the following 
typical example of our numerical study
with two heterogeneous distributed application frameworks ($n=1,2$)
having resource demands per unit workload:
\be
d_{1,1} = 5,~ d_{1,2}=1,~ d_{2,1}=1,~ d_{2,2}=5; \label{illus-d}
\ee
and two heterogeneous servers ($i=1,2$) having two different
resources with capacities:
\be
c_{1,1}=100,~ c_{1,2}=30,~ c_{2,1}=30,~ c_{2,2}=100. \label{illus-c}
\ee 
For DRF and TSF, the servers $i$
are chosen in round-robin fashion,  where the server order is  randomly
permuted in each round; DRF under such randomized round-robin (RRR) 
server selection is
the default Mesos scheduler, \cf next section. 
One can also formulate PS-DSF under RRR wherein RRR selects the server
and the PS-DSF criterion only selects the framework for that server.
Frameworks $n$ are chosen by
progressive filling with integer-valued tasking ($x$), \ie whole
tasks are scheduled.

Numerical results for scheduled workloads
for this illustrative example are given in 
Tables \ref{table:illus} \& \ref{table:illus_stddev}, and
unused resources are given in 
Tables \ref{table:unused} and \ref{table:unused_stddev}.
200 trials were performed for DRF, TSF and PS-DSF under RRR
server selection, so using
Table \ref{table:illus_stddev} we can obtain confidence intervals
for the averaged quantities given in Table \ref{table:illus}
for schedulers under RRR. For example,  the 95\% confidence interval
for task allocation of the first framework on the second server 
(\ie $(n,i)=(1,2)$) 
under TSF is 
$$(6.5 -2 \cdot 0.46/\sqrt{200},6.5 +2\cdot 0.46/\sqrt{200}) 
= (6.43,6.57).$$
Note how PS-DSF's performance under RRR is comparable to 
when frameworks and servers are jointly selected \cite{Jalal17a},
and with low variance in allocations.
We also found that RRR-rPS-DSF performed just as
rPS-DSF over 200 trials.

\begin{table}
\begin{center}
\begin{tabular}{|c||c|c|c|c||c|} 
\hline
\backslashbox{sched.}{$(n,i)$} &(1,1)&(1,2)&(2,1)&(2,2) & total \\ \hline\hline
DRF \cite{DRF,BLi15} & 6.55 & 4.69 & 4.69 & 6.55  & 22.48\\ \hline
TSF \cite{BLi16b} &  6.5 & 4.7  & 4.7 & 6.5 & 22.4\\ \hline
RRR-PS-DSF  & 19.44 & 1.15 & 1.07 & 19.42 & 41.08 \\ \hline 
BF-DRF \cite{BLi15} & 20 & 2 & 0 & 19 & 41\\ \hline
PS-DSF \cite{Jalal17a} & 19 & 0 & 2 & 20 & 41 \\ \hline
rPS-DSF  & 19 & 2 & 2 & 19 & 42 \\ \hline 
\end{tabular}
\caption{Workload allocations $x_{n,i}$ for different schedulers
under progressive filling for illustrative
example with parameters (\ref{illus-d}) and
(\ref{illus-c}). Averaged values over 200 trials reported for
the first three schedulers operating under RRR
server selection.}\label{table:illus}
\end{center}
\end{table}

\begin{table}
\begin{center}
\begin{tabular}{|c||c|c|c|c|} 
\hline
\backslashbox{sched.}{$(n,i)$} &(1,1)&(1,2)&(2,1)&(2,2)   \\ \hline\hline
DRF \cite{DRF,BLi15} & 2.31 & 0.46 & 0.46 & 2.31 \\ \hline
TSF \cite{BLi16b} &  2.29 & 0.46 &  0.46 & 2.29 \\ \hline
RRR-PS-DSF        & 0.59 & 0.99 & 1 & 0.49   \\ \hline 
\end{tabular}
\caption{Sample standard deviation 
of allocations $x_{n,i}$ for different schedulers
under RRR server selection with. Averaged  values over 200
trials reported.}\label{table:illus_stddev}
\end{center}
\end{table}

\begin{table}
\begin{center}
\begin{tabular}{|c||c|c|c|c|} 
\hline
\backslashbox{sched.}{$(i,r)$} &(1,1)&(1,2)&(2,1)&(2,2)  \\ \hline\hline
DRF \cite{BLi15} & 62.56  & 0 & 0  &62.56  \\ \hline
TSF \cite{BLi16b} &  62.8 & 0 &  0 & 62.8 \\ \hline
RRR-PS-DSF & 1.8 & 4.6 & 4.86 & 1.92  \\ \hline 
BF-DRF \cite{BLi15} & 0 & 10 & 1 &  3 \\ \hline
PS-DSF \cite{Jalal17a} & 3 & 1  & 10 & 0 \\ \hline
rPS-DSF  & 3 & 1 & 1 & 3 \\ \hline 
\end{tabular}
\caption{Unused capacities
$c_{i,r}-\sum_n x_{n,i}d_{n,r}$ for different schedulers
under progressive filling for illustrative
example with parameters (\ref{illus-d}) and
(\ref{illus-c}). Averaged values over 200 trials reported
under RRR server selection.}\label{table:unused}
\end{center}
\end{table}

\begin{table}
\begin{center}
\begin{tabular}{|c||c|c|c|c|} 
\hline
\backslashbox{sched.}{$(i,r)$} &(1,1)&(1,2)&(2,1)&(2,2)  \\ \hline\hline
DRF \cite{DRF,BLi15} & 11.09& 0 & 0  & 11.09   \\ \hline
TSF \cite{BLi16b} & 10.99  & 0 & 0   & 10.99  \\ \hline
RRR-PS-DSF        & 0.59 & 0.99 & 1 & 0.49  \\ \hline 
\end{tabular}
\caption{Sample standard deviation 
of unused capacities $c_{i,r}-\sum_n x_{n,i}d_{n,r}$ for different schedulers
under RRR server selection over 200 trials.}\label{table:unused_stddev}
\end{center}
\end{table}


We found task efficiencies improve using
{\em residual} forms of the fairness criterion.
For example, the residual  PS-DSF (rPS-DSF) criterion is
\beqa
\tK_{n,j,\ux_j} & = & x_n \max_r \frac{d_{n,r}}
{\phi_n( c_{j,r}-\sum_{n'} x_{n',j}d_{n',r})}
\eeqa
That is, this criterion makes scheduling decisions by 
progressive filling using {\em current  residual}
(unreserved) capacities based on the {\em current} allocations $x$.
From Table \ref{table:illus}, we see the improvement is
modest for the case of PS-DSF.

Improvements are also obtained by {\em best-fit} 
server selection. For example, best-fit DRF  (BF-DRF) 
first selects framework $n$ by DRF and then 
selects the server whose residual
capacity most closely matches their
resource demands $\{d_{n,r}\}_r$ \cite{BLi15}.

\section{Online experiments with Mesos}

The execution traces presented in the figures are typical of 
the multiple trials we ran.

\subsection{Introduction including background on Mesos}

The Mesos master 
(including its resource allocator, see \cite{mesos-spark-figure}) 
works in dynamic/online environment with churn in
the distributed computing/application frameworks it manages.
When all or part of a Mesos agent\footnote{an agent 
is a.k.a. server, slave or worker and is typically a virtual machine}
 becomes available, a framework is selected by Mesos and a resource allocation for it is performed. The framework accepts the offered allocation in whole or part. 
When a framework's tasks are completed, the Mesos master may be notified that the corresponding resources of the agents are released, and then 
the master will make new allocation decisions to existing or new frameworks.
Newly arrived frameworks with no allocations are given priority. 
We consider two implementations of fair resource scheduling algorithms in Mesos.

In {\bf oblivious}\footnote{called ``coarse grain" in Mesos.} allocation, the allocator is not aware of the resource demands of the frameworks\footnote{Indeed, the frameworks themselves may not be aware.}.  
A framework running an uncharacterized application may be configured to accept all resources offered to it.

In {\bf workload-characterized} 
allocation, each active framework $n$  simply
informs the Mesos allocator  of its resource demands per task,
$\{d_{n,r}\}_r$. 
The Mesos allocator selects a framework and allocates a single task worth of resources from a given agent with unassigned (released) resources.

In the following,
we compare different scheduling algorithms implemented as the Mesos allocator.
Given a pool of agents with unused resources,  
PS-DSF \cite{Jalal17a}, rPS-DSF and best-fit (BF) \cite{BLi15} 
allocations will depend on particular agents. 
When a Mesos framework (Spark job) completes, its resources from different
agents are released.
We have observed that at times the Mesos allocator sequentially schedules 
agents with available resources 
({\em i.e.,} the agents are released according to some order),
while at other times the released agents are scheduled as a pool so
that the agent-selection mechanism would be relevant.
Initially, the agents are always scheduled by the Mesos allocator as a pool.

\ifTR
In our Mesos implementation, the workflow of these two different allocations is shown in Figure \ref{flowchart}.

\begin{figure}[ht!]
	\centering
	\includegraphics[width=1\textwidth]{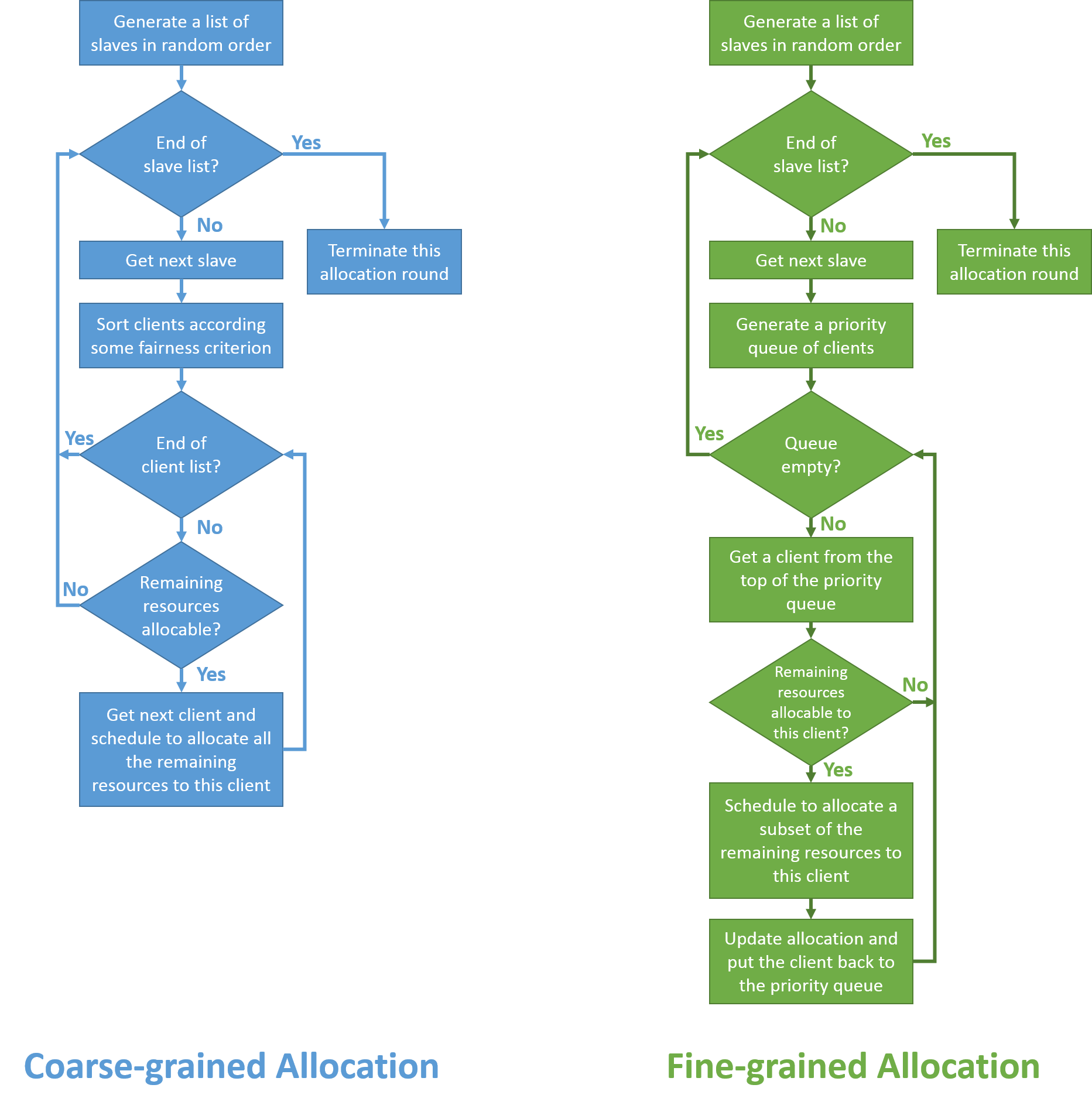}
	\caption{Flowchart of Coarse-grained/Oblivious and Fine-grained/Workload-Characterized Allocation.}
	\label{flowchart}
\end{figure}

\fi

\subsection{Running Spark on Mesos}

In our experiments, the frameworks will operate under the distributed programming platform Spark in client mode. 
Each Spark job (Mesos framework) is divided into multiple tasks (threads).
Multiple Spark {\em executors} will be allocated for a Spark job. 
The executors can simultaneously run a certain maximum number of tasks
depending on how many cores on the executor and how many cores are
required per task; when a task completes, the executor informs the driver to request
another, {\em i.e.} executors pull in work.
Each executor is a Mesos task in the default ``coarse-grained" mode
\cite{mesos-spark}
and an executor 
resides in a container of a Mesos agent \cite{mesos-container}. Plural executors can 
simultaneously reside on a single Mesos agent.
An executor usually terminates as the entire Spark job terminates 
\cite{spark-dynamic-allocation}. 
When starting a Spark job, the resources 
required to start an executor ($d$) and the maximum number of executors that can be created to execute the tasks of the  job, may be specified.
The Spark driver 
will attempt to use as much of its allocated resources as possible.

In a typical configuration,
Spark employs three classical parallel-computing techniques:
jobs are divided into microtasks (typically based on fine
partition of the dataset on which work is performed);
when underbooked, executors pull work  (tasks)
from a driver; and the driver employs time-out at program 
	barriers\footnote{where parallel executed tasks {\em all} need to 
complete before the program can proceed} to detect
straggler tasks and relaunch them on new executors (speculative execution) \cite{spark-config}. 
In this way, Spark can reduce (synchronization) delays
at barriers while not needing to know either 
the execution speed of the executors nor the 
resources required to achieve a particular execution time
of the tasks. On the other hand, microtasking does incur
potentially significant overhead compared to an approach
with larger tasks whose resource 
needs have been better characterized, {\em i.e.}, as $\{d_{n,r}\}$
resources per task\footnote{what may be called ``coarse grain" in the context
of Spark.}. 

\subsection{Experiment Configuration}

In our experiments, there are two Spark submission groups (``roles" in Mesos' jargon): group Pi submits jobs that accurately calculate $\pi =3.1415...$ via Monte Carlo simulation;  group WordCount submits word-count jobs for a 700MB+ document. The executors of Pi require 2 CPUs and about 2 GB memory (Pi is CPU bottlenecked), while those of WordCount require 1 CPU and about 3.5 GB memory (WordCount is memory bottlenecked). Each group has five job submission queues, which means there could be ten jobs running on the cluster at the same time. Each queue initially has fifty jobs. Again, each job is divided into tasks and tasks are run in plural Spark executors (Mesos tasks)
running on different Mesos agents.

The Mesos agents run on six servers (AWS c3.2xlarge virtual-machine instances), two each of three types in our cluster. A type-1 server provides 4 CPUs and 14 GB memory, so it 
would be well utilized by 4 WordCount tasks. A type-2 server provides 8 CPUs and 8 GB memory, so it would be well utilized by 4 Pi tasks. A type-3 server provides 6 CPUs and 11 GB memory, so it would be well utilized by 2 Pi and 2 WordCount tasks. 
The Mesos master operates in a c3.2xlarge with 8 cores and 15 GB memory.

\ifTR
The experiment setup is illustrated in Figure \ref{set-up}.

\begin{figure}[ht!]
	\centering
	\includegraphics[width=1\textwidth]{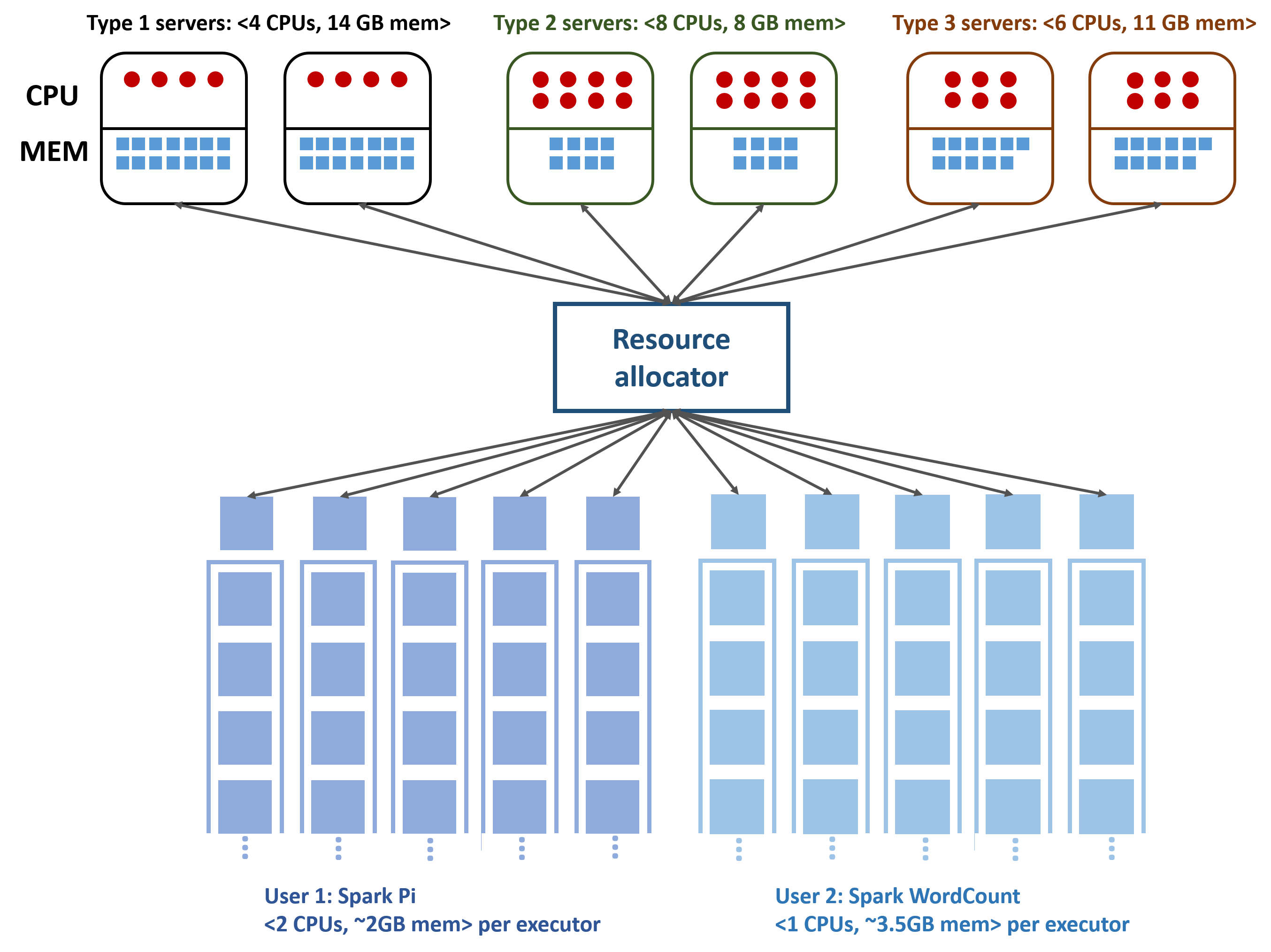}
	\caption{Experiment setup.}
	\label{set-up}
\end{figure}

\fi

\subsection{Prototype implementation}

We modified the allocator module of Mesos (version 1.5.0) to use different scheduling criteria;
in particular, criteria   depending on the agent/server so that agents are
not necessarily selected in RRR fashion when a pool of them is available.
We also modified the driver in Spark to pass on a framework $n$'s resource needs per task ($\{d_{n,r}\}$) in workload-characterized mode.
\ifanon
Our code is open-source available.
\else
Our code is available here \cite{mesos-code,spark-code}.
\fi

\subsection{Experimental Results for Different Schedulers}
We ran the same total workload for the four different Mesos allocators all
under Randomized Round-Robin (RRR) agent selection: 
oblivious DRF (Mesos default), oblivious PS-DSF, workload-characterized DRF, and workload-characterized PS-DSF.  (In this section, we drop the
``RRR" qualifier).
A summary of our results is that overall execution time 
is improved under workload characterization
and under allocations that are agent/server specific.

\subsubsection{DRF vs. PS-DSF in oblivious mode}

\begin{figure}[ht!]
	\centering
	\ifTR
	\includegraphics[width=0.7\textwidth]{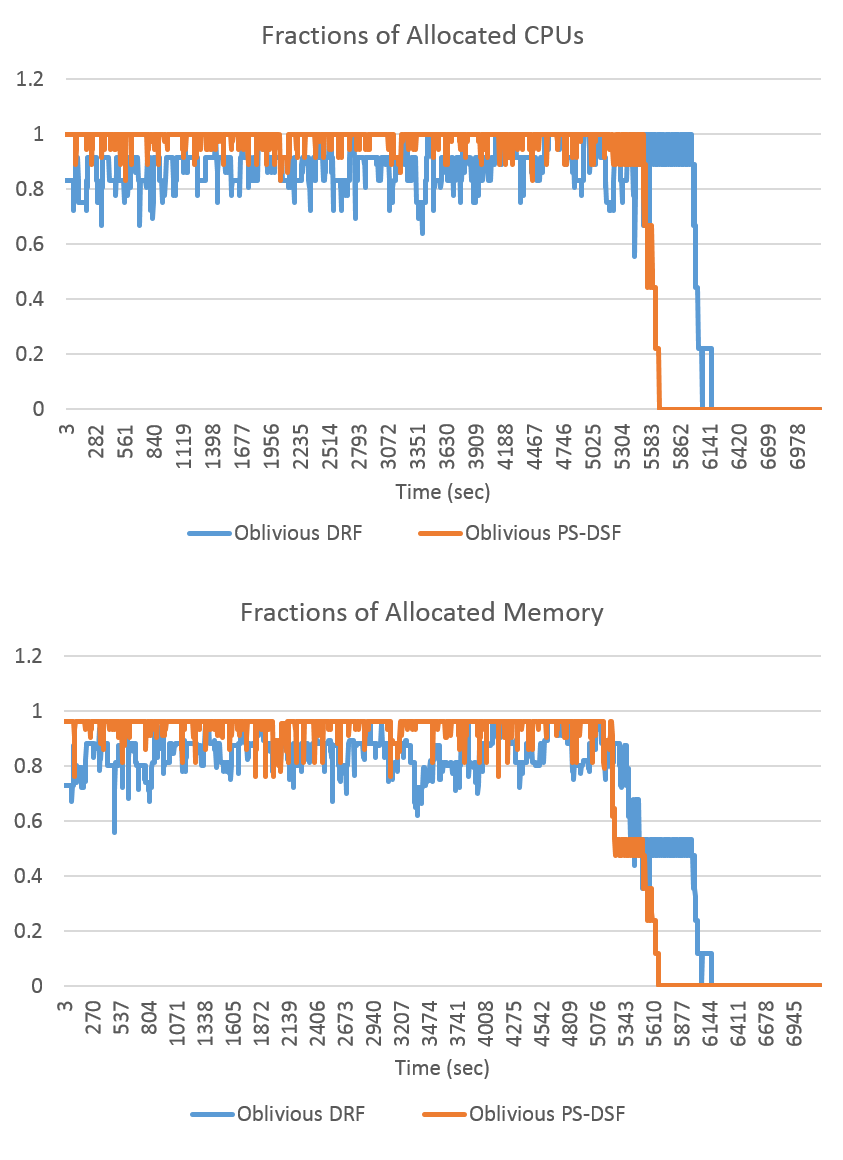}
	\fi
	\caption{Comparison between DRF and PS-DSF in oblivious mode.}
	\label{coarse_drf_psdsf}
\end{figure}

The resource allocation under different fairness criteria are shown in Figure \ref{coarse_drf_psdsf}. It can be seen that PS-DSF can achieve higher resource utilization than DRF because it ``packs" tasks better into heterogeneous servers. As a result, the entire job-batch under PS-DSF finishes earlier.
Also note that at the end of the experiment, there is a sudden drop in allocated memory percentage. This is because the memory-intensive Spark WordCount jobs finish earlier and CPU is the bottleneck resource for the remaining Spark Pi jobs. 

\subsubsection{Schedulers in workload-characterized mode}
\begin{figure}[ht!]
	\centering
	\ifTR
	\includegraphics[width=0.7\textwidth]{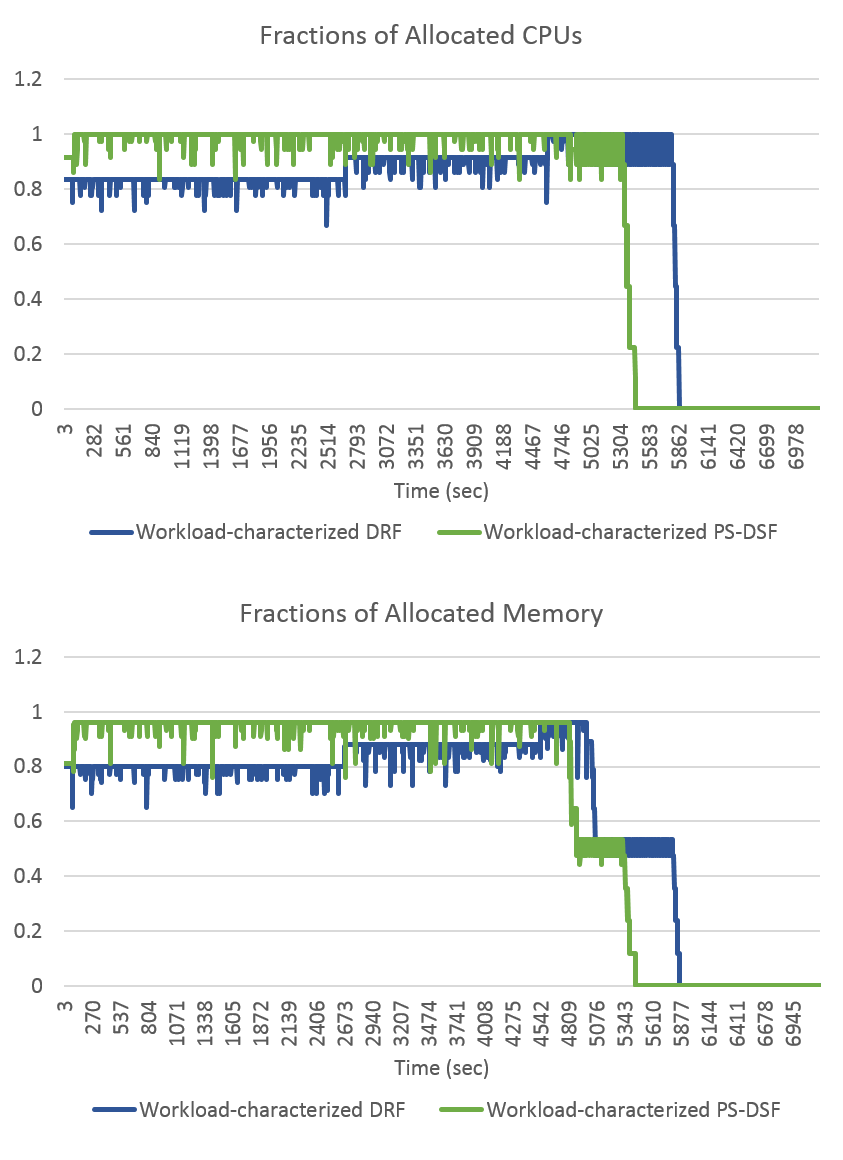}
	\fi
	\caption{Comparison between DRF and PS-DSF in workload-characterized mode.}
	\label{fine_drf_psdsf}
\end{figure}

The experimental results under workload-characterized mode, as shown in Fig. \ref{fine_drf_psdsf}, are consistent with their oblivious counterparts - PS-DSF has higher resource utilization than DRF. 
Also note that the resource utilizations in workload-characterized mode have less variance than those in oblivious mode, which will be explained in Sec. \ref{sec:coarse_vs_fine}.

\begin{figure}[ht!]
	\centering
	\ifTR
	\includegraphics[width=0.7\textwidth]{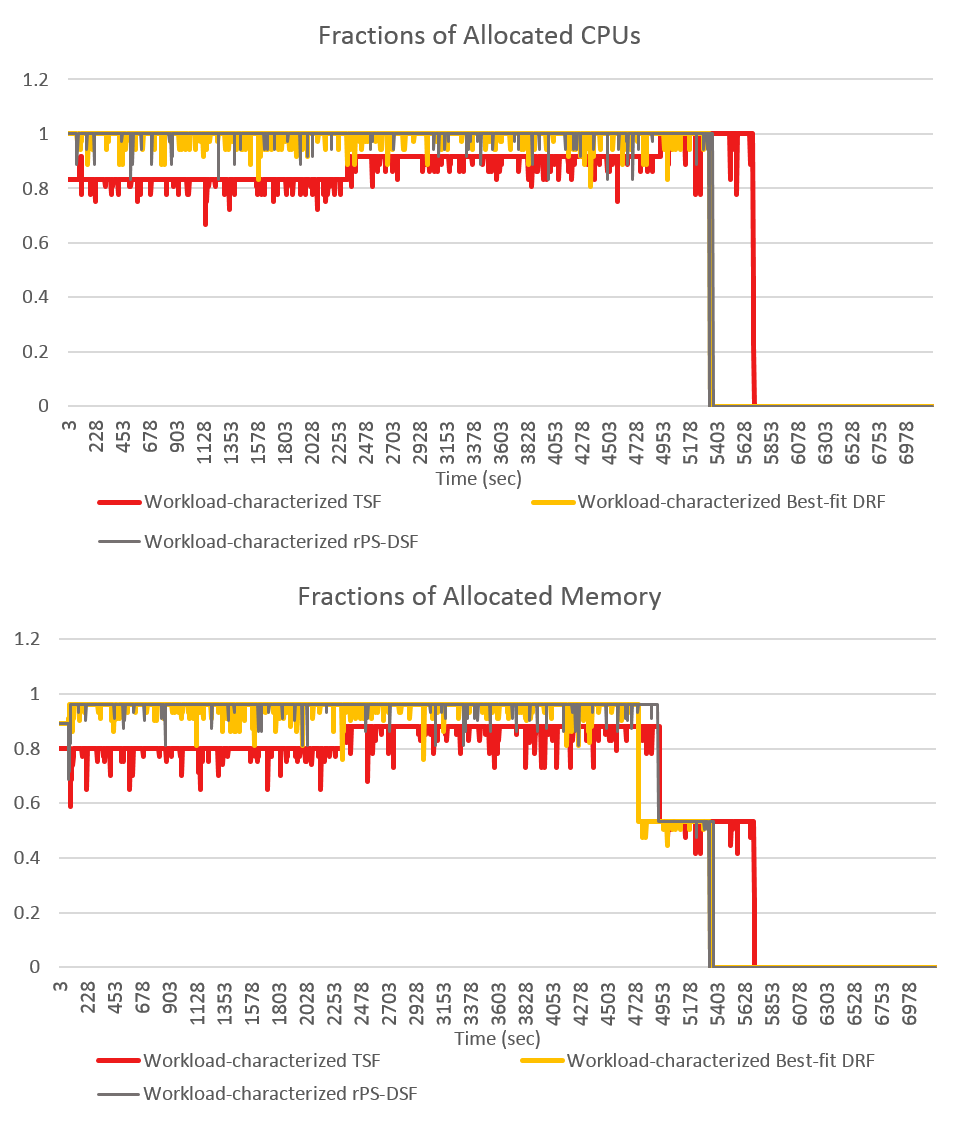}
	\fi
	\caption{Comparison among TSF, Best-fit DRF and rPS-DSF (in the workload characterized mode).}
	\label{tsf_bfdrf_rpsdsf}
\end{figure}

In Figure \ref{tsf_bfdrf_rpsdsf}, we compare 
TSF \cite{BLi16b} under RRR\footnote{Note that \cite{BLi16b}
also describes experimental results for a Mesos implementation of TSF.}, 
rPS-DSF (under RRR),
and BF-DRF (again, ``best fit" is an agent-selection mechanism when
there is a pool of agents to choose from). 
From the figure, the execution times of BF-DRF and -rPS-DSF 
are comparable  to PS-DRF
(but \cf Section \ref{BF-DRF-v-RRR-rPS-DSF}) and
shorter than TSF  (which is comparable to DRF).

\subsubsection{Oblivious versus Workload Characterized modes}\label{sec:coarse_vs_fine}

\begin{figure}[ht!]
	\centering
	\ifTR
	\includegraphics[width=0.7\textwidth]{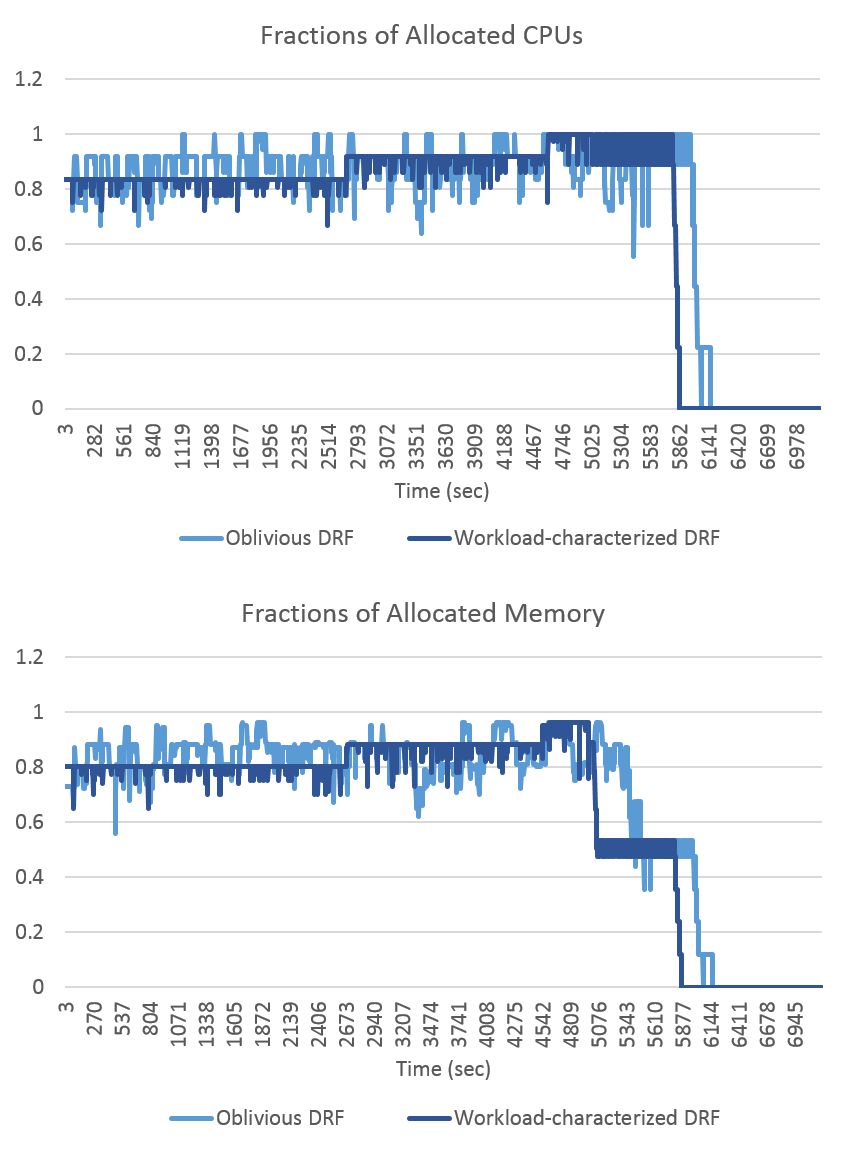}
	\fi
	\caption{Comparison between oblivious and workload-characterized modes under DRF.}
	\label{drf_coarse_fine}
\end{figure}

\begin{figure}[ht!]
	\centering
	\ifTR
	\includegraphics[width=0.7\textwidth]{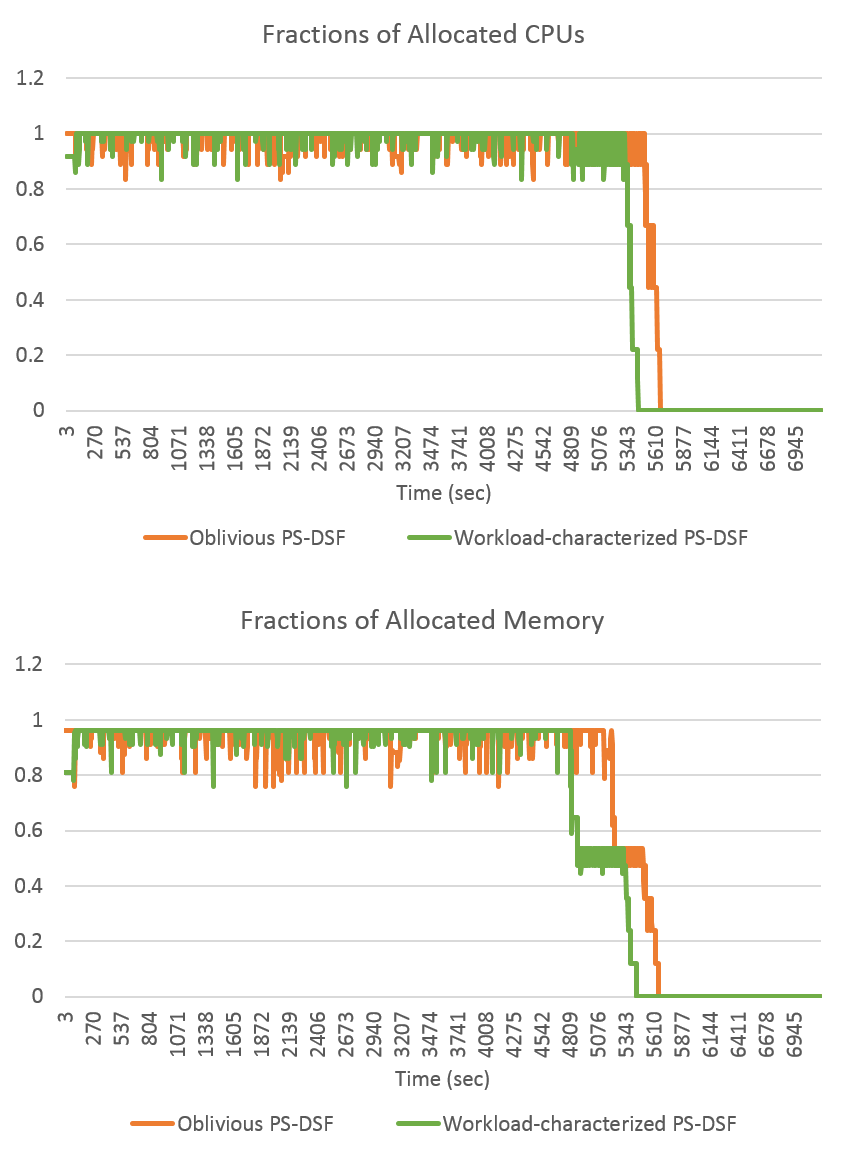}
	\fi
	\caption{Comparison between oblivious and workload-characterized modes under PS-DSF.}
	\label{psdsf_coarse_fine}
\end{figure}

We also compared oblivious and workload-characterized allocation for the same scheduling algorithm. 
Again, when a Spark job finishes, its executors may
not simultaneously release resources from the Mesos allocator's point-of-view.
So under oblivious allocation, it's possible that multiple Spark frameworks can share the same server, as is typically the case under workload-characterized scheduling.
However, oblivious allocation is a coarse-grained enforcement of progressive filling, where the resources are less evenly distributed among the frameworks - some frameworks may receive the entire remaining resources on a agent in a single offer, leaving nothing available for others.
From Figures \ref{drf_coarse_fine}-\ref{psdsf_coarse_fine},  note how 
under oblivious allocation the amount of allocated resources drops more sharply when a Spark job ends, and variance of utilized resources under oblivious allocation is larger than under workload-characterized. Consequently, the entire job-batch tends to finish sooner under workload-characterized allocator, as we see in Figures \ref{drf_coarse_fine}-\ref{psdsf_coarse_fine}.

\ifTR
\subsection{With Homogeneous Servers}
We also did experiments in a cluster with six type-3 servers (6 CPUs, 11 GB memory). In Figure \ref{fine_drf_psdsf_hm} we show that DRF and PS-DSF have nearly identical performance with homogeneous servers.

\begin{figure}[ht!]
	\centering
	\ifTR
	\includegraphics[width=0.7\textwidth]{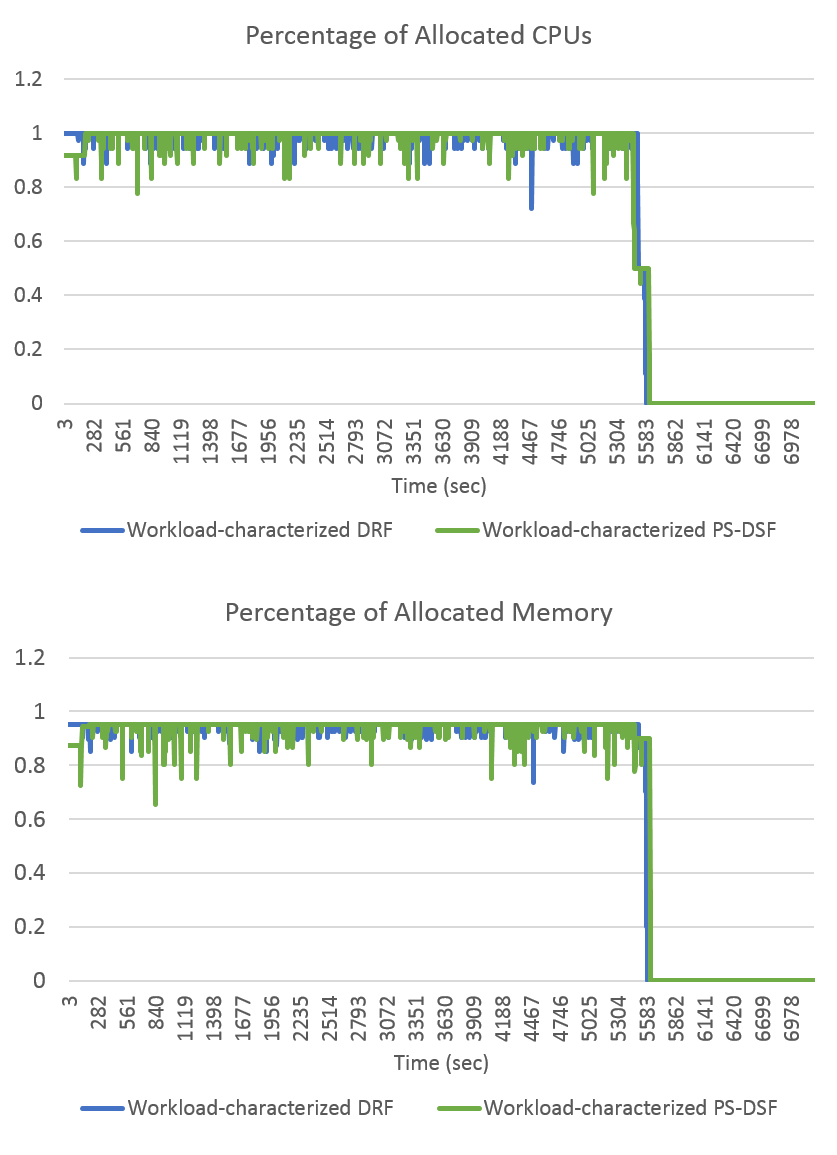}
	\else
	\includegraphics[width=0.7\columnwidth]{./spark-exp/fine_drf_psdsf_hm.png}
	\fi
	\caption{Workload-characterized DRF and PS-DSF with homogeneous servers.}
	\label{fine_drf_psdsf_hm}
\end{figure}
\fi

\subsection{BF-DRF versus rPS-DSF}\label{BF-DRF-v-RRR-rPS-DSF}

Finally, with a different experimental set-up,
we compare BF-DRF (which first selects the framework and then
selects the ``best fit" from among available agents/servers) and
a representative of a family of server-specific schedulers,
rPS-DRF under RRR.
Consider a case where there are three servers, one of each of the above server types (types 1-3).  

Suppose under a current allocation, we have one Spark-Pi and two Spark-WordCount executors on the type-1 server, two Spark-Pi and one Spark-WordCount executors on the type-2 server, and two Spark-Pi and two Spark-WordCount executors on the type-3 server. So, whenever a Pi or WordCount framework releases its executor's resources back to the cluster, its DRF ``score" is reduced so the scheduler will always sends a resource offer to the same client framework in this scenario.  On the other hand, rPS-DSF will make a decision considering the amount of (remaining) resources on the server, and so will make a more efficient allocation.

\begin{figure}[ht!]
	\centering
	\ifTR
	\includegraphics[width=0.7\textwidth]{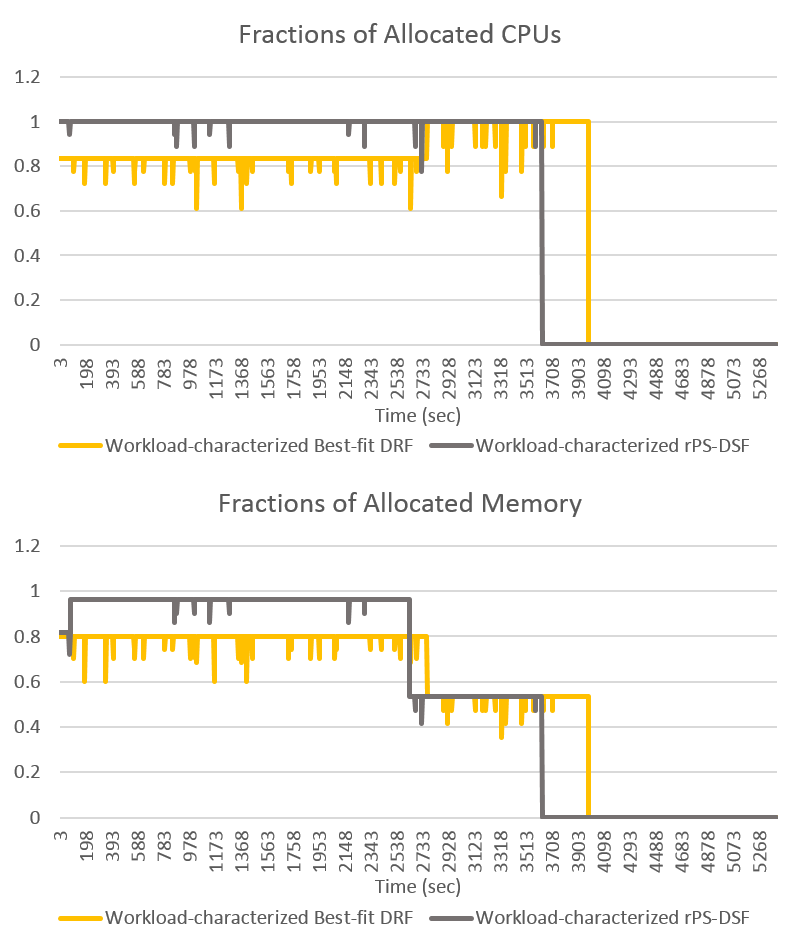}
	\fi
	\caption{Performance of Best-fit DRF and rPS-DSF given initial suboptimal allocation.}
	\label{slave_aware}
\end{figure}

We illustrate this with the example of Figure \ref{slave_aware}. In this experiment, we let each group submit their Spark jobs through five queues with 20 jobs each. To create the above scenario, instead of exposing all the servers to the client frameworks, we register servers one by one from type-1 to type-3. From the figure, note that both rPS-DSF and BF-DRF have an initiall inefficient memory allocation, but rPS-DSF is able to adapt and quickly increase its memory efficiency, while BF-DRF does not.

\bibliographystyle{plain}
\bibliography{../../../latex/scheduling}

\begin{thebibliography}{10}

\bibitem{DRF}
A.~Ghodsi, M.~Zaharia, B.~Hindman, A.~Konwinski, S.~Shenker, and I.~Stoica.
\newblock Dominant resource fairness: Fair allocation of multiple resource
  types.
\newblock In {\em Proc. USENIX NSDI}, 2011.

\bibitem{Jalal17a}
J.~Khamse-Ashari, I.~Lambadaris, G.~Kesidis, B.~Urgaonkar, and Y.Q. Zhao.
\newblock {Per-Server Dominant-Share Fairness (PS-DSF): A Multi-Resource Fair
  Allocation Mechanism for Heterogeneous Servers}.
\newblock In {\em Proc. IEEE ICC, Paris}, May 2017.

\bibitem{mesos-container}
{Apache Mesos - Containerizers}.
\newblock http://mesos.apache.org/documentation/
  latest/containerizer-internals/.

\bibitem{mesos-spark-figure}
{Apache Mesos - Mesos Architecture}.
\newblock http://mesos.apache.org/documentation/latest/architecture/.

\bibitem{mesos-code}
Mesos multi-scheduler.
\newblock https://github.com/yuquanshan/mesos/tree/multi-scheduler.

\bibitem{spark-dynamic-allocation}
{Apache Spark - Dynamic Resource Allocation}.
\newblock https://spark.apache.org/docs/latest/job-scheduling.html.

\bibitem{mesos-spark}
{Apache Spark - Running Spark on Mesos}.
\newblock https://spark.apache.org/docs/latest/running-on-mesos.html.

\bibitem{spark-config}
{Apache Spark - Spark Configuration}.
\newblock https://spark.apache.org/docs/latest/configuration.html.

\bibitem{spark-code}
Spark with resource demand vectors.
\newblock https://github.com/yuquanshan/spark/tree/d-vector.

\bibitem{BLi16b}
W.~Wang, B.~Li, B.~Liang, and J.~Li.
\newblock Multi-resource fair sharing for datacenter jobs with placement
  constraints.
\newblock In {\em Proc. Supercomputing}, Salt Lake City, Utah, 2016.

\bibitem{BLi15}
W.~Wang, B.~Liang, and B.~Li.
\newblock Multi-resource fair allocation in heterogeneous cloud computing
  systems.
\newblock {\em IEEE Transactions on Parallel and Distributed Systems},
  26(10):2822--2835, Oct. 2015.

\end{thebibliography}

\end{document}